\documentclass[%
reprint,
 amsmath,amssymb,
 aps,
]{revtex4-2}

\usepackage{subfigure}
\usepackage{float}
\usepackage{slashed}
\usepackage{xcolor}
\usepackage{amsmath}
\usepackage{subcaption}
\usepackage[english]{babel}
\usepackage{lineno, blindtext}
\usepackage{tikz}
\usepackage{adjustbox}
\usepackage{rotating}
\usepackage{rotfloat}
\usepackage{verbatim}
\usepackage[utf8]{inputenc}
\usepackage{graphicx}% Include figure files
\usepackage{dcolumn}% Align table columns on decimal point
\usepackage{bm}% bold math
\usepackage[super]{nth}
\usepackage{lipsum}
\newcommand\Tdiag[4]{%
    \multicolumn{1}{|p{#2}|}{\hskip-\tabcolsep
    \begin{tikzpicture}[%
                baseline={(0,-.25\baselineskip)},
                every node/.style={outer sep=0pt,inner sep=#1}]
    \node[minimum width={#2+1\tabcolsep-\pgflinewidth},
        minimum height=2\baselineskip-\pgflinewidth+\extrarowheight,
        use as bounding box] (box) {};
    \draw[line cap=round] (box.north west) -- (box.south east);
    \node[anchor=south west,text width=.75*#2,align=left] at (box.south west) {#3};
    \node[anchor=north east,text width=.75*#2,align=right] at (box.north east) {#4};
\end{tikzpicture}\hskip-\tabcolsep}}

\usepackage{hyperref}
\usepackage{xcolor}
\hypersetup{
  colorlinks   = true, %Colours links
  urlcolor     = red, %Colour for external hyperlinks
  linkcolor    = blue, %Colour of internal links
  citecolor   = blue %Colour of citations
}

\usepackage{threeparttable}
\begin{document}

\preprint{APS/123-QED}

\title{Search for the production of dark matter in the framework of Mono-Z$^{\prime}$ portal at the ILC simulated electron-positron collisions at $\sqrt{s} = 500$ GeV}

\author{S. Elgammal}
 \altaffiliation[sherif.elgammal@bue.edu.eg]{}%Lines break automatically or can be forced with \\
\affiliation{%
Centre for theoretical physics, The British University in Egypt. %\textbackslash\textbackslash
}%


\date{\today}% It is always \today, today,

\begin{abstract}
{In the present work, we study the possible production of the light neutral gauge boson (Z$^{\prime}$) candidates, which originated from a simplified-model scenario based on 
the Mono-Z$^{\prime}$ model, in association with dark matter.
This study has been performed by studying events with dimuon plus missing transverse energy produced in the simulated electron-positron collisions at the foreseen International Linear Collider (ILC), operating at 500 GeV center of mass energy and integrated luminosity of 1000 fb$^{-1}$.
In case no new physics has been discovered, we set upper limits at 95\% confidence level 
on the masses of various particles in the model as, spin-1 (Z$^{\prime}$), as well as the fermionic dark matter.}

\vspace{0.75cm}
%\hrule
%\begin{description}
%\item[Keywords]
%Dark matter, New gauge boson, The Large Hadron Collider LHC, The Compact Muon Solenoid CMS
%\end{description}
\end{abstract}

\maketitle

%\tableofcontents

%=============================================================================

%\begin{acknowledgments}
%We wish to acknowledge the support of the author community in using
%REV\TeX{}, offering suggestions and encouragement, testing new versions,
%\end{acknowledgments}

%\nocite{*}

%================================================

\section{Introduction}
\label{sec:intro}
\begin{comment}
Several cosmological evidences, based on recent observations \cite{planck2015, planck2018, bullet_cluster}, have confirmed the existence of an unknown form of matter named the Dark Matter (DM), which occupies around 27\% of the total energy distribution in the Universe.
Various types of candidates have been suggested to compose DM \cite{R9}. One of the accepted hypotheses, which is in an agreement with the observed density, is that the bulk of DM has the form of electrically-neutral, non-baryonic, non-decaying, and weakly-interacting massive particles (WIMPs) \cite{R6}. 
\end{comment}
The Standard Model of particle physics (SM) has succeeded in describing many features of 
nature that we observe in our experiments \cite{SMcource}. 
The most famous example, arguably, is the agreement between the SM prediction and the experimental
measurement of the electroweak theory, which led to the old discovery of $W^{\pm}$, $Z^{0}$ bosons and the recent discovery of the SM Higgs boson by ATLAS experiment \cite{higgsDiscoveryATLAS} and CMS experiment \cite{higgsDiscoveryCMS}.

Although its success, SM is well known that it is not a complete theory. 
It describes only the visible, baryonic, matter in the Universe, which represents about 
5\% of the energy density distribution content of the Universe.
The rest is attributed to the mysterious dark energy and dark matter (DM). 
\begin{comment}
As it does not contain a description of the fourth known fundamental force gravity. In addition to that, the particles and forces contained within the SM describe only the visible, baryonic, matter in the Universe. Cosmological observations have proven that this baryonic matter forms only around 5\% of the energy density distribution content of the Universe, with the rest attributed to the mysterious dark energy and dark matter (DM). 
\end{comment}
Many cosmological observations, based on recent observations by \cite{planck2015, planck2018, bullet_cluster}, support the existence of dark matter.
These observations suggest that dark matter is non-decaying and weakly interacting massive particles. %\cite{R6}.
So, SM could be considered as a low-energy manifestation of other theories realized at high energy, generically known as BSM (Beyond the Standard Model) theories \cite{SMandBSM}. 

Both the ATLAS and CMS Collaborations have previously searched for the massive extra neutral gauge boson Z$^{\prime}$, which is induced by the Grand Unified Theory (GUT) and Supersymmetry \cite{Extra-Gauge-bosons, gaugeboson1, LR-symmetry11, Super-symmetry12}, with no evidence of their existence using the full RUN II period of the LHC data \cite{zprime, zprimeATLAS}.
Results by the CMS have excluded, at 95\% confidence level (CL), the existence of Z$^{\prime}$ in the mass range between 0.6 - 5.15 TeV, while the ATLAS experiment has excluded the mass range between 0.6 - 5.1 TeV. 

In addition, many searches for DM have been performed by analyzing the data collected by 
the CMS and ATLAS experiments during RUN II. These searches rely on the production of a visible object "X", which recoils against the large missing transverse energy from the dark matter particles leaving a signature of ($\text{X} + E^{\text{miss}}_{T}$) in the detector \cite{R38}. 
The visible particle could be an SM particle like W, Z bosons or jets \cite{R35, atlasMonoZ}, photon \cite{photon, photonATLAS} or SM Higgs boson \cite{R36, monoHiggsAtlas1, monoHiggsAtlas2}. 

The foreseen International Linear Collider (ILC) is an electron-positron collider, which has been proposed to operate at 500 GeV center of mass energy ($\sqrt{s}$) as a startup with possible upgrade to $\sqrt{s} = 1000$ GeV in run II \cite{ilc1,ilc2,ilc3,ilc4}.
The linear electron-positron colliders are characterized by controllable initial particle energy, low QCD background compared to hadron colliders, and controllable 
beam polarization.

In this analysis, we present a search for light neutral gauge bosons (Z$^{\prime}$) (i.e. $M_{Z^{\prime}} < 90$ GeV), which are originated from a simplified-model scenario, which 
is known as the light vector (LV), in the framework of Mono-Z$^{\prime}$ model \cite{R1}, 
at the ILC simulated electron-positron collisions with 500 GeV center of mass energy.  
The topology of the studied simulated events is dimuon, from the decay of Z$^{\prime}$, plus large missing transverse energy which is attributed to dark matter.
A similar analysis has been presented for searching for dark matter in association with the visible particle being a $Z$ boson decaying to dimuon at the ILC \cite{R45055}. 

ATLAS experiment has performed a search for dark matter, in the framework of Mono-Z$^{\prime}$ model, with the muonic decay channel of Z$^{\prime}$ at the LHC \cite{atlasMonoZprime}. 
This search has excluded Z$^{\prime}$ masses between 200 and 1000 GeV, and seated separate limits on the coupling of the Z$^{\prime}$ to the SM leptons, $\texttt{g}_{l}$. 
Thus, $\texttt{g}_{l}$ ranges from 0.01–0.025 and 0.02–0.38 for Z$^{\prime}$ masses between 200 and 1000 GeV are excluded for the light and heavy dark-sector in the light-vector scenario. 
A search, which has been done by CMS experiment in final states with dimuon, excluded Z$^{\prime}$ boson in the mass region below 200 GeV \cite{lowMasszprime}. 
In addition, the LHCb has placed tight constraints on the Z$^{\prime}$ mass and Z$^{\prime}$-dimuon coupling in the dimuon mass region that the current paper probes \cite{lowMasszprimeLHCb1, lowMasszprimeLHCb2}.
%- see e.g. 2007.03923, 1710.02867.

If the Z$^{\prime}$ does not couple to quarks, the HL-LHC and future hadron colliders would provide no limits on the existence of Z$^{\prime}$. Then the future electron-positron colliders as ILC will play an important role. 

In this paper, we will first present the theoretical framework of the simplified model based on the Mono-Z$^{\prime}$ portal and its free parameters in section \ref{section:model}. Next, we will explain the simulation techniques used for generating events for both the signal and the SM background samples in section \ref{section:MCandDat}. We will then discuss the selection cuts and the analysis strategy in section \ref{section:AnSelection}. Finally, we will present the results and provide a summary of this analysis in sections \ref{section:Results} and \ref{section:Summary}, respectively.

\begin{figure} [h!]
\centering
\resizebox*{6.0cm}{!}{\includegraphics{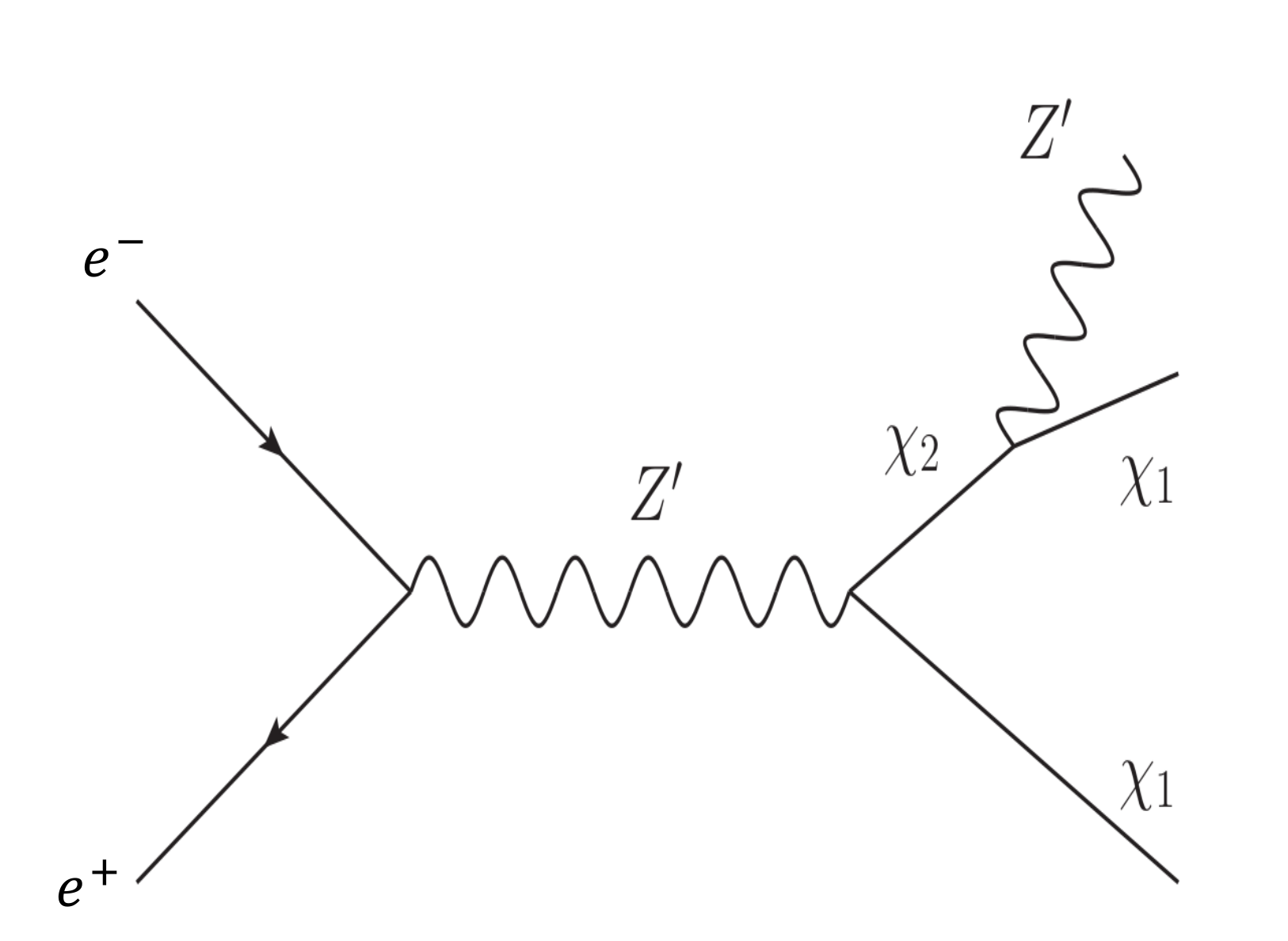}}
\caption{Feynman diagram for the Light Vector (LV) scenario based on Mono-Z$^{\prime}$ model; for the production of neutral light gauge boson (Z$^{\prime}$) in association 
to dark matter ($\chi_{1}$) pair \cite{R1}.}
\label{figure:fig1}
\end{figure}

%=================================================
\section{The simplified model in the framework of Mono-Z$^{\prime}$ portal}
\label{section:model}
The Mono-Z$^{\prime}$ model, discussed in \cite{R1}, assumes that dark matter is produced from electron-positron collisions at the ILC via a new light gauge boson Z$^{\prime}$. In the Mono-Z$^{\prime}$ model, the dark matter can be produced through three different scenarios. However, we will focus only on the Light Vector (LV) scenario, which is presented in figure \ref{figure:fig1} for the current analysis.

The proposed dark matter can be produced through the process of pair annihilation of electron-positron mediated by a light vector boson Z$^{\prime}$, which then undergoes two dark matters, a light dark matter $(\chi_{1})$ and a heavy one $(\chi_{2})$. 
$\chi_{2}$ is heavy enough to decay to a Z$^{\prime}$ and another light dark matter $\chi_{1}$ (i.e. $\chi_{2}$ $\rightarrow$ Z$^{\prime}$ $\chi_{1}$) as shown in figure \ref{figure:fig1}.

The interaction term, in the Lagrangian, between the dark fermions and Z$^{\prime}$ is given by \cite{R1} 
\begin{equation}
    \frac{\texttt{g}_{DM}}{2} Z^{\prime}_{\mu}\large(\bar{\chi_{2}}\gamma^{\mu}\gamma^{5}\chi_{1} + \bar{\chi_{1}}\gamma^{\mu}\gamma^{5}\chi_{2}\large), \nonumber
\end{equation}
where $\texttt{g}_{DM}$ is the coupling of Z$^{\prime}$ to the dark matter $\chi_{1}$ and $\chi_{2}$. In the rest of the paper, the notation $(\chi)$ refers to $(\chi_{1})$.
For future reference, the coupling of the Z$^{\prime}$ to the visible leptons will be represented by $\texttt{g}_{l}$, while the coupling of the Z$^{\prime}$ to the dark matter will be represented by $\texttt{g}_{DM}$ in the rest of this paper.  
%============================================================
\begin{table} [h!]    
\centering
\begin {tabular} {ll}
\hline
\hline
Scenario & \hspace{3pt} Masses assumptions \\
\hline
\\
    & \hspace{4pt} $M_{\chi_{1}} = 1, 5, ...,200$ GeV \\
Light dark sector & \\
    & \hspace{4pt} $M_{\chi_{2}} = M_{\chi_{1}} + M_{Z'} + 25$ GeV
  \\
%\hline
%\\
%& \hspace{4pt} $M_{\chi_{1}} = M_{Z'} / 2$ GeV \\
%Heavy dark sector & \\
% &\hspace{4pt} $M_{\chi_{2}} = 2M_{Z'}$ GeV
 \\
\hline
\hline
\end {tabular}
\vspace{3pt}
\caption{The Light mass assumptions for the dark sector for the light vector scenario \cite{R1}.}
\label{table:tab1}
\end{table}
%=========================================================

The only allowed decays in the LV scenario are assumed to be the decay of  ${Z}'\rightarrow\chi_{1}\chi_{2}$, $\chi_{2}\rightarrow {Z'}\chi_{1}$ and ${Z}'\rightarrow\mu^{+}\mu^{-}$. 
Where the total decay width of the ${Z}'$ and ${\chi_{2}}$ can be calculated given the values of the masses of ${Z}'$ and dark matter and the coupling constants.

The free parameters in this scenario are the lightest dark matter mass $M_{\chi_{1}}$,
$M_{\chi_{2}}$, the Z$^{\prime}$ mass $(M_{Z^{\prime}})$ and the coupling of Z$^{\prime}$ to both leptons and dark matter particles $\texttt{g}_{l}$ and $\texttt{g}_{DM}$ respectively.

The CMS and ATLAS detectors have conducted intensive searches for Z$^{\prime}$ over many years. As a result, it has been proven that heavy neutral gauge bosons (Z$^{\prime}$) do not exist in the mass range between 0.2 to 5.15 TeV. Therefore, we focus our investigation on the production of light neutral gauge bosons (Z$^{\prime}$) in the mass range below 90 GeV at the ILC.

To achieve this, the LV scenario has been considered with the use of the light-dark sector case for acquiring mass to dark matter particles ($\chi_{1}$ and $\chi_{2}$) as presented in table \ref{table:tab1}. This is a specific choice made to fix $M_{\chi_{1}}$ and $M_{\chi_{2}}$ according to a prescription given in \cite{R1}. It is important to note that this is only one of the possible choices.

The ATLAS and CMS experiments have established strict limitations on the value of $\texttt{g}_{l}$ at low $M_{Z^{\prime}}$ based on the 4-muon final states, neutrino trident production, and LEPII constraints. Two publications (\cite{R37} and \cite{R777}) demonstrate the existence of these limitations. For the range of $M_{Z^{\prime}}$ values between 10 and 80 GeV that the present paper investigates, $\texttt{g}_{l}$ needs to be approximately 0.003. According to \cite{R1}, $\texttt{g}_{DM}$ is equal to 1.0. 
While the values of the masses ($M_{Z^{\prime}}$, $M_{\chi_{1}}$ and $M_{\chi_{2}}$) are not fixed but are scanned over.

The signal of the processes being studied is identified by two leptons with opposite charges. These leptons are created from the decay of Z$^{\prime}$, and are accompanied by a large amount of missing transverse energy caused by the stable dark matter $\chi_{1}$. Therefore, the events we are examining can be described as having the topology of ($\mu^{+}\mu^{-} + E^{miss}_{T}$).

\begin{table*}
\centering
%\begin{flushleft}
%\tiny
%\scriptsize
\fontsize{9.pt}{12pt}
\selectfont
\begin{tabular}{|c|c|c|c|c|c|c|c|c|c|}
\hline
\Tdiag{.4em}{2.1cm}{$M_{\chi_{1}}$}{$M_{Z'}$}  & 10 & 20 & 30 & 40 & 50 & 60 & 70 & 80 & 90  \\
\hline
%\hline
1  & 
    $7.73\times10^{-1}$ & 
    $7.08\times10^{-1}$ & 
    $6.73\times10^{-1}$ & 
    $6.49\times10^{-1}$ & 
    $6.28\times10^{-1}$ & 
    $6.12\times10^{-1}$ & 
    $6.01\times10^{-1}$ & 
    $5.88\times10^{-1}$ &
    $5.78\times10^{-1}$ \\

\hline
5  & 
    $7.63\times10^{-1}$ & 
    $6.87\times10^{-1}$ & 
    $6.52\times10^{-1}$ & 
    $6.29\times10^{-1}$ & 
    $6.09\times10^{-1}$ & 
    $5.95\times10^{-1}$ & 
    $5.82\times10^{-1}$ & 
    $5.70\times10^{-1}$ &
    $5.60\times10^{-1}$ \\
    
\hline
10  & 
    $7.62\times10^{-1}$ & 
    $6.65\times10^{-1}$ & 
    $6.30\times10^{-1}$ & 
    $6.05\times10^{-1}$ & 
    $5.86\times10^{-1}$ & 
    $5.73\times10^{-1}$ & 
    $5.59\times10^{-1}$ & 
    $5.48\times10^{-1}$ &
    $5.39\times10^{-1}$ \\
    \hline
25  & 
    $6.44\times10^{-1}$ & 
    $6.20\times10^{-1}$ & 
    $5.75\times10^{-1}$ & 
    $5.47\times10^{-1}$ & 
    $5.27\times10^{-1}$ & 
    $5.12\times10^{-1}$ & 
    $4.98\times10^{-1}$ & 
    $4.86\times10^{-1}$ &
    $4.76\times10^{-1}$ \\
    \hline
50  & 
    $6.35\times10^{-1}$ & 
    $5.57\times10^{-1}$ & 
    $4.92\times10^{-1}$ & 
    $4.59\times10^{-1}$ & 
    $4.37\times10^{-1}$ & 
    $4.19\times10^{-1}$ & 
    $4.03\times10^{-1}$ & 
    $3.90\times10^{-1}$ &
    $3.77\times10^{-1}$ \\

\hline
100 & 
    $6.30\times10^{-1}$ & 
    $3.80\times10^{-1}$ & 
    $3.15\times10^{-1}$ & 
    $2.82\times10^{-1}$ & 
    $2.59\times10^{-1}$ & 
    $2.40\times10^{-1}$ & 
    $2.24\times10^{-1}$ & 
    $2.10\times10^{-1}$ &
    $1.94\times10^{-1}$ \\
\hline
125 & 
    $4.46\times10^{-1}$ & 
    $2.69\times10^{-1}$ & 
    $2.20\times10^{-1}$ & 
    $1.93\times10^{-1}$ & 
    $1.73\times10^{-1}$ & 
    $1.56\times10^{-1}$ & 
    $1.41\times10^{-1}$ & 
    $1.28\times10^{-1}$ &
    $1.17\times10^{-1}$ \\
\hline
150 &  
    $1.85\times10^{-1}$ & 
    $1.62\times10^{-1}$ & 
    $1.31\times10^{-1}$ & 
    $1.11\times10^{-1}$ & 
    $9.57\times10^{-2}$ & 
    $8.26\times10^{-2}$ & 
    $7.07\times10^{-2}$ & 
    $5.97\times10^{-2}$ &
    $4.95\times10^{-2}$ \\
\hline
170 &  
    $1.35\times10^{-1}$ & 
    $8.95\times10^{-2}$ & 
    $7.07\times10^{-2}$ & 
    $5.77\times10^{-2}$ & 
    $4.67\times10^{-2}$ & 
    $3.72\times10^{-2}$ & 
    $2.86\times10^{-2}$ & 
    $2.12\times10^{-2}$ &
    $1.48\times10^{-2}$ \\

    \hline
175 &  
    $1.10\times10^{-1}$ & 
    $7.41\times10^{-2}$ & 
    $5.81\times10^{-2}$ & 
    $4.64\times10^{-2}$ & 
    $3.68\times10^{-2}$ & 
    $2.82\times10^{-2}$ & 
    $2.08\times10^{-2}$ & 
    $1.45\times10^{-2}$ &
    $9.22\times10^{-3}$ \\
\hline
200 & 
    $2.48\times10^{-2}$ & 
    $1.84\times10^{-2}$ & 
    $1.28\times10^{-2}$ & 
    $8.08\times10^{-3}$ & 
    $4.50\times10^{-3}$ & 
    $2.02\times10^{-3}$ & 
    $6.27\times10^{-4}$ & 
    $1.15\times10^{-4}$ &
    $8.19\times10^{-6}$ \\
    \hline
\end {tabular}
\caption{The light vector scenario production cross sections times branching ratios (in fb) at Leading Order (LO) for different choices of the DM mass $M_{\chi_{1}}$ (in GeV) and Z$^{\prime}$ mass $M_{Z^{\prime}}$ (in GeV); for the light-dark sector mass assumption with the following couplings constants $\texttt{g}_{l} = 0.003,~\texttt{g}_{DM} = 1.0$.
Given that the polarized degrees of electron and positron beams are $P_{e^{-}} = 0.8$, 
$P_{e^{+}} = -0.3$ at the ILC with $\sqrt{s}$ = 500 GeV.}
\label{table:tabchi}
%\end{flushleft}
\end{table*}
%=========================================================
%%%%%%%%%%%%% plots step-1 %%%%%%%%%%%%%%%%%%%%%%%%%%%%%%%
\begin{figure} %[h!]
\centering
\resizebox*{9.3cm}{!}{\includegraphics{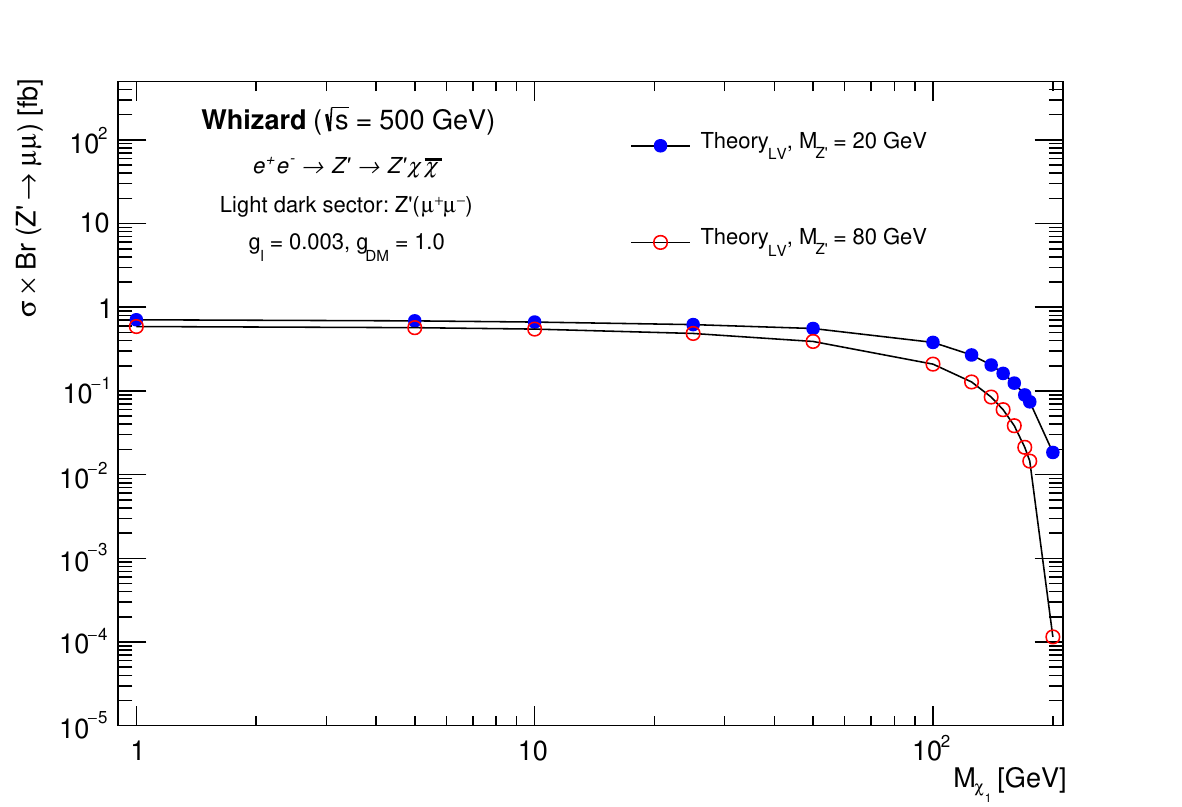}}
\caption{Dependence of cross sections for the signal process $e^{+} e^{-} \rightarrow \chi \bar{\chi} Z^{\prime}$($Z^{\prime} \rightarrow \mu^{+}\mu^{-})$ induced by the light vector (LV) on the DM mass with $\sqrt{s}$ = 500 GeV and the electron-positron beams polarization 
are $P_{e^{-}} = 0.8$, $P_{e^{+}} = -0.3$.
The blue dots refers to $M_{Z^{\prime}} = 20$ GeV, while the red dots for $M_{Z^{\prime}} = 80$ GeV.}

%\caption{The behavior of the LV scenario production cross section times branching ratios versus the dark matter mass ($M_{\chi_{1}}$), at the ILC with $\sqrt{s} = 500$ GeV, are represented by blue dots for $M_{Z^{\prime}} = 20$ GeV and in red circles for $M_{Z^{\prime}} = 80$ GeV.The electron and positron beams polarization are $P_{e^{-}} = 0.8$, $P_{e^{+}} = -0.3$.}
\label{figure:crossSec}
\end{figure}
%%%%%%%%%%%%%%%%%%%%%%%%%%%%%%%%%%%%%%%%%%%%
 
%%%%%%%%%%%%% plots step-1 %%%%%%%%%%%%%%%%%%%%%%%%%%%%%%%
\begin{figure} %[h!]
\centering
\resizebox*{9.3cm}{!}{\includegraphics{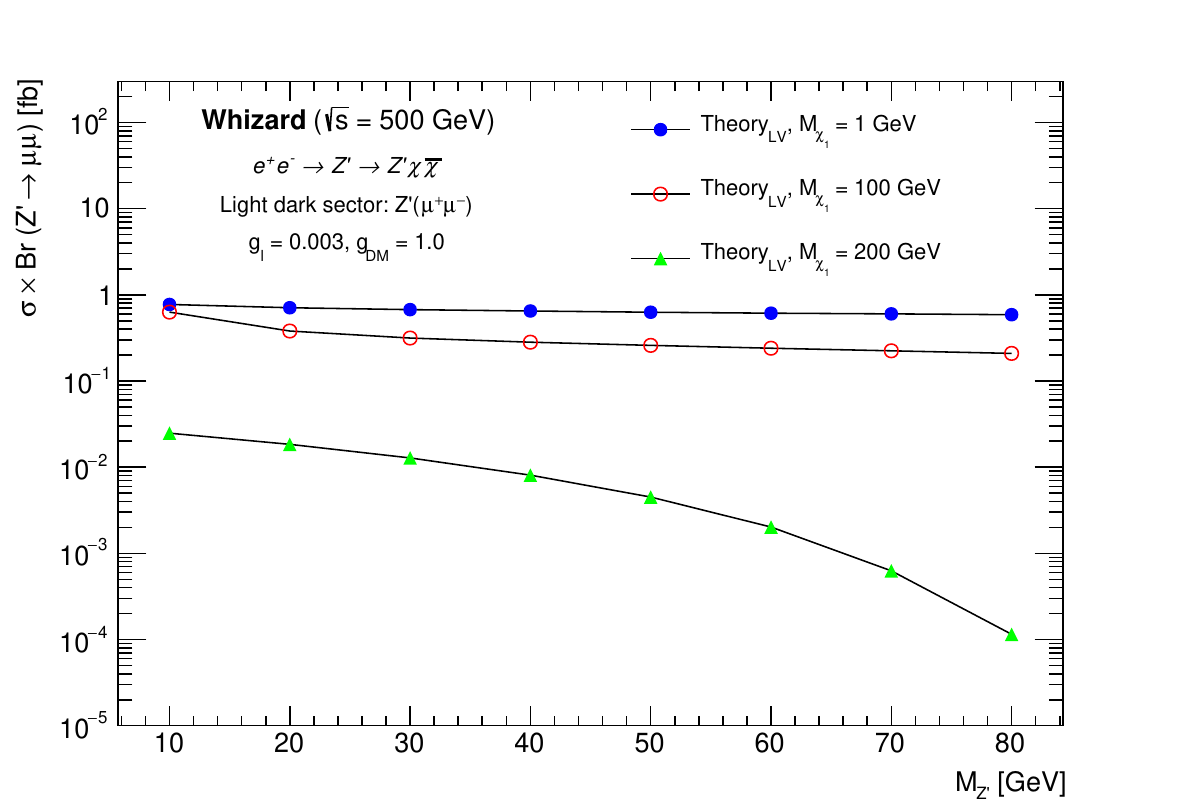}}
\caption{Dependence of cross sections for the signal process $e^{+} e^{-} \rightarrow \chi \bar{\chi} Z^{\prime}$($Z^{\prime} \rightarrow \mu^{+}\mu^{-})$ induced by the light vector (LV) on the $Z^{\prime}$ mass with $\sqrt{s}$ = 500 GeV and the electron-positron beams polarization are $P_{e^{-}} = 0.8$, $P_{e^{+}} = -0.3$.
The blue dots refers to $M_{\chi_{1}} = 1$ GeV, the red dots for $M_{\chi_{1}} = 100$ GeV and the green dots for $M_{\chi_{1}} = 200$ GeV.}
\label{figure:crossSec2}
\end{figure}

%%%%%%%%%%%%%%%%%%%%%%%%%%%%%%%%%%%%%%%%%%%%
\begin{comment}
%===============================================================
%========  sigma Vs Mchi ===========================
\begin{table}
\centering
%\begin{flushleft}
%\tiny
%\scriptsize
%\fontsize{4.7pt}{11pt}
\selectfont

\begin{tabular}{|c|c|}
\hline
$M_{Z^{\prime}}$ (GeV) & $\sigma \times \text{BR (pb)}$\\
%\hline
\hline
150  & $1.73\times10^{-2}$ \\
\hline
200  & $0.51\times10^{-2}$ \\
\hline
250 & $0.18\times10^{-2}$ \\ 
\hline
300 & $0.74\times10^{-3}$ \\ 
\hline
350 & $0.32\times10^{-3}$ \\ 
\hline
400 & $0.14\times10^{-3}$ \\
\hline
450 & $0.69\times10^{-4}$ \\ 
\hline
500 & $0.36 \times10^{-4}$ \\ 
\hline
600 & $0.11\times10^{-4}$ \\
\hline 
700 & $0.33 \times10^{-5}$\\ 
\hline
\end {tabular}
\caption{Cross section times branching ratio (in pb) for the heavy dark sector in the DF scenario calculated for different sets of the masses $M_{Z'}$, with the following couplings constants $\texttt{g}_{SM} = 0.1,~\texttt{g}_{DM} = 1.0$ and at $\sqrt{s} = 8$ TeV.}
\label{table:tabchi2}
%\end{flushleft}
\end{table}

\end{comment}
%=========================================================
%%%%\newpage
%$~~$
%\newpage
%\section{The CMS detector and reconstruction techniques}
%\label{section:CMS}

%===========================================================
\section{Simulation of signal samples and SM backgrounds}
\label{section:MCandDat}
%===================================================================
%\subsection{Backgrounds estimation and systematic uncertainties}
%\subsection{Backgrounds estimation}
%\label{section:background}
The SM background processes yielding muon pairs in the signal region are 
Drell-Yan (DY $\rightarrow \mu^+\mu^-$) production, 
the production of top quark pairs ($\text{t}\bar{\text{t}} \rightarrow \mu^+\mu^- + 2b + 2\nu$) and production of diboson 
($W^{+}W^{-} \rightarrow \mu^+\mu^- + 2\nu$,  
$ZZ \rightarrow \mu^+\mu^- + 2\nu$ and 
$ZZ \rightarrow 4\mu$).

The LV scenario signal samples and corresponding SM background processes were generated using WHIZARD event generator 3.1.1 \cite{MG5}. The ISR effect was included and interfaced with Pythia 6.24 for parton shower model and hadronization \cite{R34}. For a fast detector simulation of ILD detector, DELPHES package \cite{delphes} was used. These were generated from electron-positron collisions at the ILC with a 500 GeV center of mass energy, which corresponds to the circumstances of RUN I. The polarized degrees of electron and positron beams are $P_{e^{-}} = 0.8$ and $P_{e^{+}} = -0.3$, respectively.

For the scenario where light vectors are produced along with dark matter particles ($\chi_{1}$ and $\chi_{2}$), we have considered the mass assumptions as summarized in table \ref{table:tab1}. Assuming $\texttt{g}_{l} = 0.003$ and $\texttt{g}_{DM} = 1.0$, table \ref{table:tabchi} shows the production cross section times branching ratios at Leading Order (LO) for various mass points of Z$^{\prime}$ and DM. 
The polarized degrees of electron and positron beams are $P_{e^{-}} = 0.8$ and $P_{e^{+}} = -0.3$, respectively, at the ILC with $\sqrt{s}$ = 500 GeV.

In figure \ref{figure:crossSec}, the production cross sections multiplied by branching ratios are plotted against the mass of the dark matter ($M_{\chi_{1}}$) for two different Z$^{\prime}$ mass points ($M_{Z^{\prime}}$ = 20 and 80 GeV). These signal samples were generated at the ILC with $\sqrt{s}$ = 500 GeV, and the electron-positron beams polarization was $P_{e^{-}} = 0.8$ and $P_{e^{+}} = -0.3$, respectively. 
This graph shows that the cross sections multiplied by branching ratios remain relatively flat until the mass of the dark matter reaches 100 GeV. After that point, they decrease rapidly.

Figure \ref{figure:crossSec2} illustrates the cross sections times branching ratios against the mass of the neutral gauge boson (Z$^{\prime}$) for three different dark matter mass values: 1 GeV, 100 GeV, and 200 GeV. It has been observed that there is a significant decrease in the cross sections times branching ratios as the dark matter mass increases, while the mass of Z$^{\prime}$ also affects this trend.

The Monte Carlo simulations were used to generate the SM background samples and calculate their corresponding cross-sections for this analysis. The calculations were done in leading order and can be found in table \ref{table:tab3}. 
The signal samples and SM background processes were estimated from these simulations and were normalized to their respective cross sections and an integrated luminosity of 1000 fb$^{-1}$.

An ad-hoc flat 10\% uncertainty is applied to cover all possible systematic effects.

%===========================================================
%======= MCS data Sets ===================================
\begin{table*} 
\centering
\begin {tabular} {|l|l|l|c|l|}
\hline
Process \hspace{1cm} & Deacy channel  & Generator  & {$\sigma \times \text{BR} ~(\text{fb})$} & Order \\
\hline
\hline
$\text{DY}$ & $\mu^{+}\mu^{-}$ & Whizard & 1767 & LO\hspace{6cm}\\
\hline
$\text{t}\bar{\text{t}}$ & $\mu^{+}\mu^{-} + 2b + 2\nu$ & Whizard& 10.4 & LO \\
\hline
WW & $\mu^{+}\mu^{-} + 2\nu$ & Whizard & 232.8 & LO \\
\hline
ZZ & $\mu^{+}\mu^{-} + 2\nu$  & Whizard & 3.7 & LO \\
\hline
ZZ  & $4\mu$ & Whizard & 0.5 & LO \\
\hline
\end {tabular}
\vspace{3pt}
\caption{The simulated SM backgrounds generated from electron-positron collisions at the ILC 
with the polarized degrees of electron and positron beams are $P_{e^{-}} = 0.8$, $P_{e^{+}} = 
-0.3$ at $\sqrt{s} = 500$ GeV. Their corresponding cross-section times branching ratio for each process and the generation order are presented. 
Names of these MC samples and the used generators are stated as well.}
\label{table:tab3}
\end{table*}
%===========================================================

%=========================================================
%==============================================================
\section{Event selection}
\label{section:AnSelection}
The event selection process has been designed to reconstruct a final state consisting of two muons with low transverse momentum $(p_{T})$, along with missing transverse energy accounting for the dark matter candidate. The selection is made by applying cuts on various kinematic parameters. 

Both muons must pass a preliminary selection that includes the following criteria: \\
- $p^{\mu}_{T}$ (GeV) $> 10$,\\
- $|\eta^{\mu}|$ (rad) $<$ 2.5,\\
- $\text{IsolationVar} < 0.1$. 

Here, $\text{"IsolationVar"}$ represents the isolation cut in DELPHES software to reject muons produced inside jets. This cut requires that the scalar $p_{T}$ sum of all muon tracks within a cone of $\Delta R = 0.5$ around the muon candidate, excluding the muon candidate itself, should not exceed 10\% of the $p_{T}$ of the muon. 

Therefore, each event is selected with two opposite charge muons.

%%%%%%%%%%%%% plots step-1 %%%%%%%%%%%%%%%%%%%%%%%%%%%%%%%
\begin{figure} [h!]
\centering
\resizebox*{9cm}{!}{\includegraphics{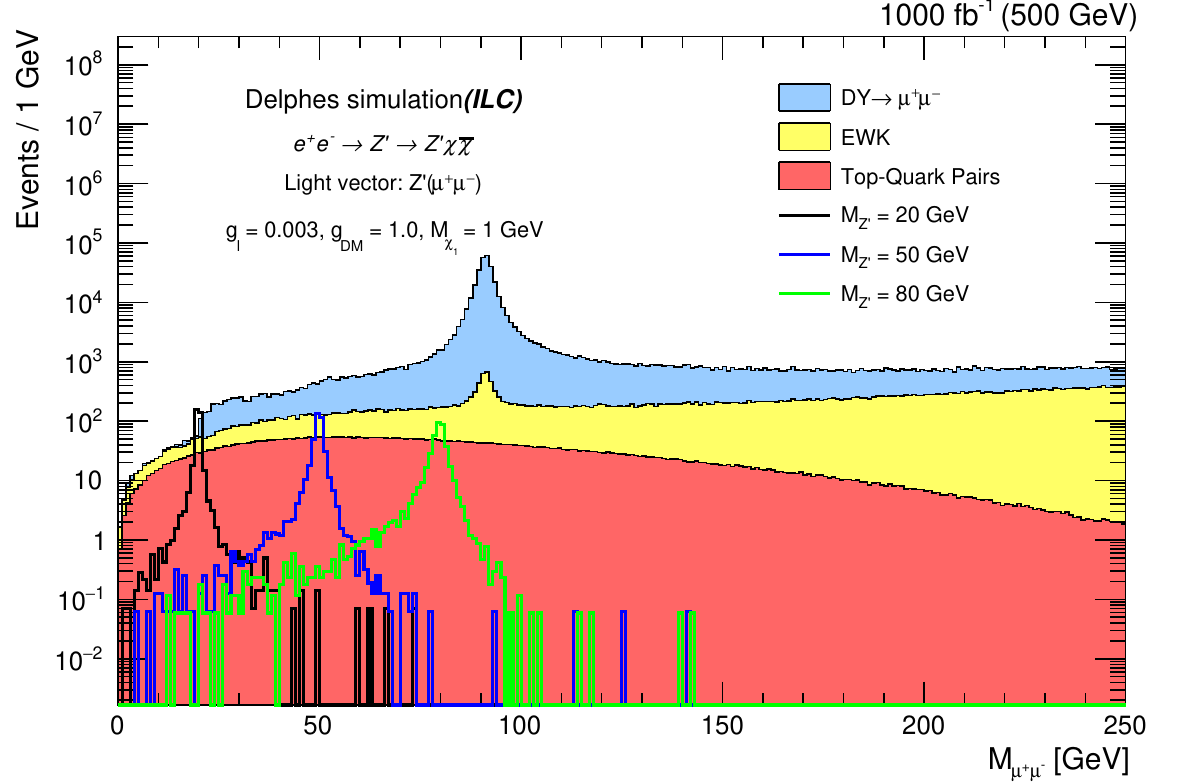}}
\caption{The measured dimuon invariant mass spectrum, after applying preliminary selection(i)  summarized in table \ref{cuts}, for the estimated SM backgrounds and different choices of neutral gauge boson (Z$^{\prime}$) masses generated based on the LV scenario, with dark matter mass ($M_{\chi_{1}} = 1$ GeV).}
\label{figure:fig3}
\end{figure}
%%%%%%%%%%%%%%%%%%%%%%%%%%%%%%%%%%%%%%%%%%%%
\begin{figure} [h!]
\centering
\resizebox*{9cm}{!}{\includegraphics{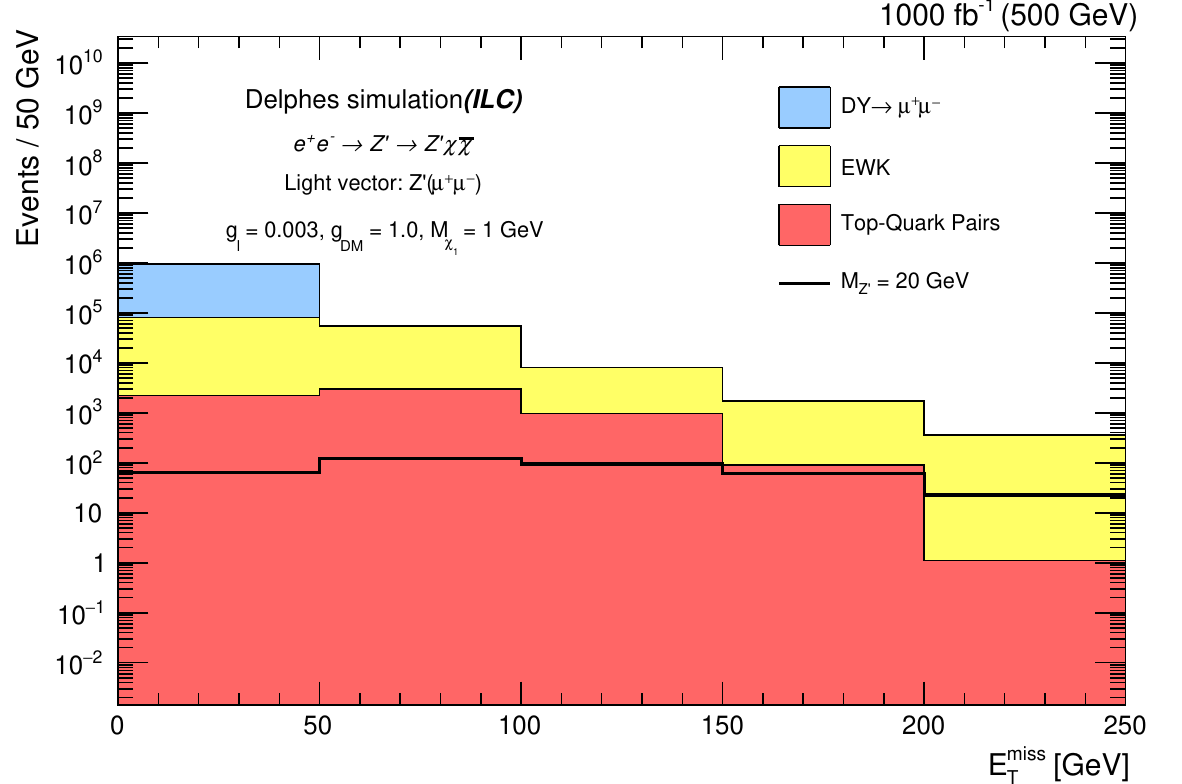}} 
%\resizebox*{10cm}{!}{\includegraphics{.png}}\hspac{}
  \caption{The distribution of the missing transverse energy, after applying selection(i) summarized in table \ref{cuts}; for the expected SM backgrounds, and mass of $Z^{\prime}$ ($M_{Z^{\prime}}$ = 20 GeV) produced by the LV scenario, with mass of dark matter ($M_{\chi_{1}} = 1$ GeV).}
\label{figure:fig4}
\end{figure}
%===================================================================

Figure \ref{figure:fig3} shows the distribution of the dimuon invariant mass; 
the cyan histogram represents the Drell-Yan background, the yellow histogram stands for the vector boson pair backgrounds (WW and ZZ), and the red histogram represents the $t\bar{t}$ background. 
These histograms are stacked. 
While the signals of the LV scenario, which have been generated with different masses of the Z$^{\prime}$ boson and fixing the dark matter mass ($M_{\chi_{1}} = 1$ GeV), are represented by different colored lines, and are overlaid. 

Figure \ref{figure:fig4} presents the corresponding distributions of the missing transverse energy for a signal sample with $M_{Z^{\prime}} = 20$ GeV and $M_{\chi_{1}} = 1$ GeV, as well as for SM backgrounds. The figure indicates that the signal sample is heavily contaminated with backgrounds across the entire missing transverse energy range. Therefore, it is crucial to implement more stringent criteria to distinguish the signals from SM backgrounds, as explained in the following paragraph.
%=====================================================================
%%%%%%%%%%%%% event selection %%%%%%%%%%%%%%%%%%%%%%%%%%%%%%%
%============================================================
%\begin{comment}
%==== Final selection table ================
\begin{table*}
    \centering
    \begin{tabular}{|c|c|c|}
   % \begin {tabular} {| l \space |c|c|}
\hline
Step & Variable & Requirements \\
\hline
    \hline
                & $p^{\mu}_{T}$ (GeV) & $>$ 10 \\
Selection(i)    & $|\eta^{\mu}|$ (rad) & $<$ 2.5 \\
                & $\Sigma_{i} p^{i}_{T}/p^{\mu}_{T}$ & $<$ 0.1 \\
    
\hline
 %&&\\
   & Mass window (GeV) & $0.9 \times M_{Z^{\prime}} < M_{\mu^{+}\mu^{-}} < M_{Z^{\prime}} + 25$ \\
    % &&\\
     &$\Delta\phi_{\mu^{+}\mu^{-},\vec{E}_{T}^{miss}}$ & $> 2.9$ \\
Selection(ii)     &$|E_{T}^{\mu^{+}\mu^{-}} - E_{T}^{miss}|/E_{T}^{\mu^{+}\mu^{-}}$ & $< 0.4$ \\
     &$\Delta R(\mu^{+}\mu^{-})$ & $< 1.8$ \\
     &&\\
    \hline
    \end{tabular}
    \caption{Summary of cut-based final event selection used in the analysis.}
    \label{cuts}
\end{table*}
%\end{comment}
%============================================================
%%%%%%%%%%%%% plots step-1 %%%%%%%%%%%%%%%%%%%%%%%%%%%%%%%
\begin{figure}%[h!]
\centering
\subfigure[$|E_{T}^{\mu^{+}\mu^{-}} - E_{T}^{miss}|/E_{T}^{\mu^{+}\mu^{-}}$]{
  \includegraphics[width=92mm]{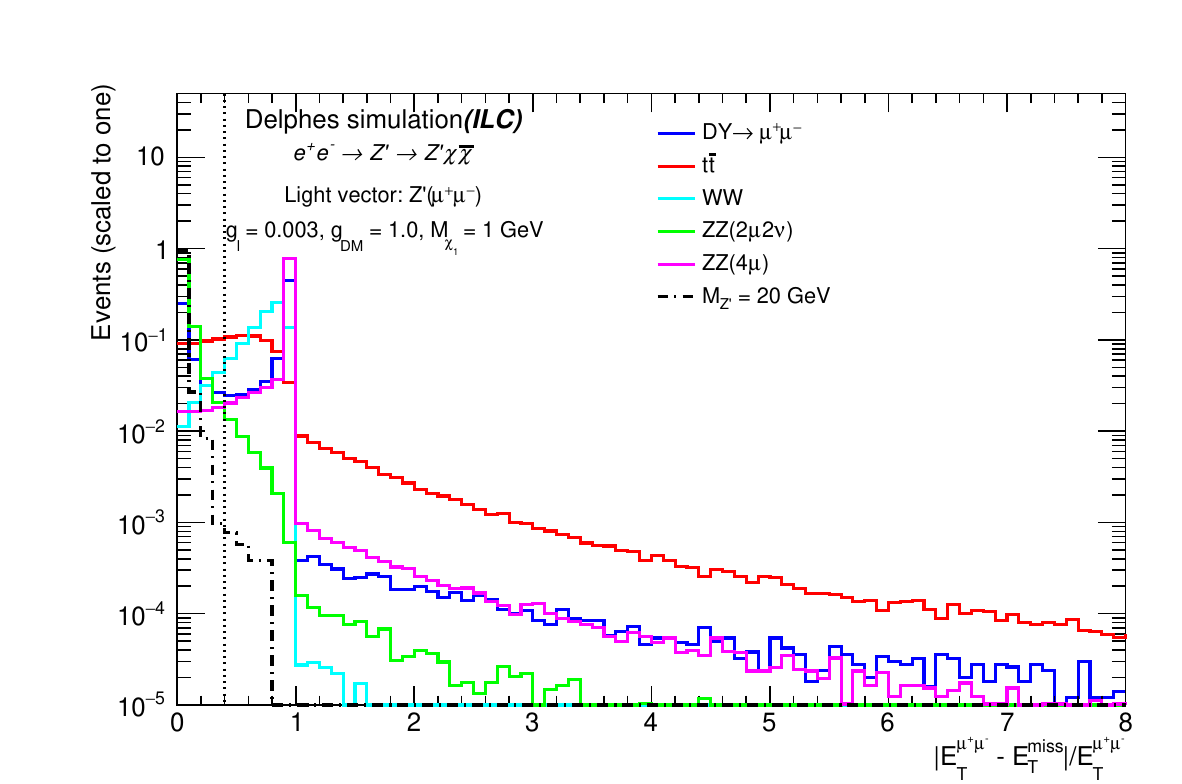}
}
\hspace{0mm}
%\centering
\subfigure[$\Delta\phi_{\mu^{+}\mu^{-},\vec{E}_{T}^{miss}}$]{
  \includegraphics[width=92mm]{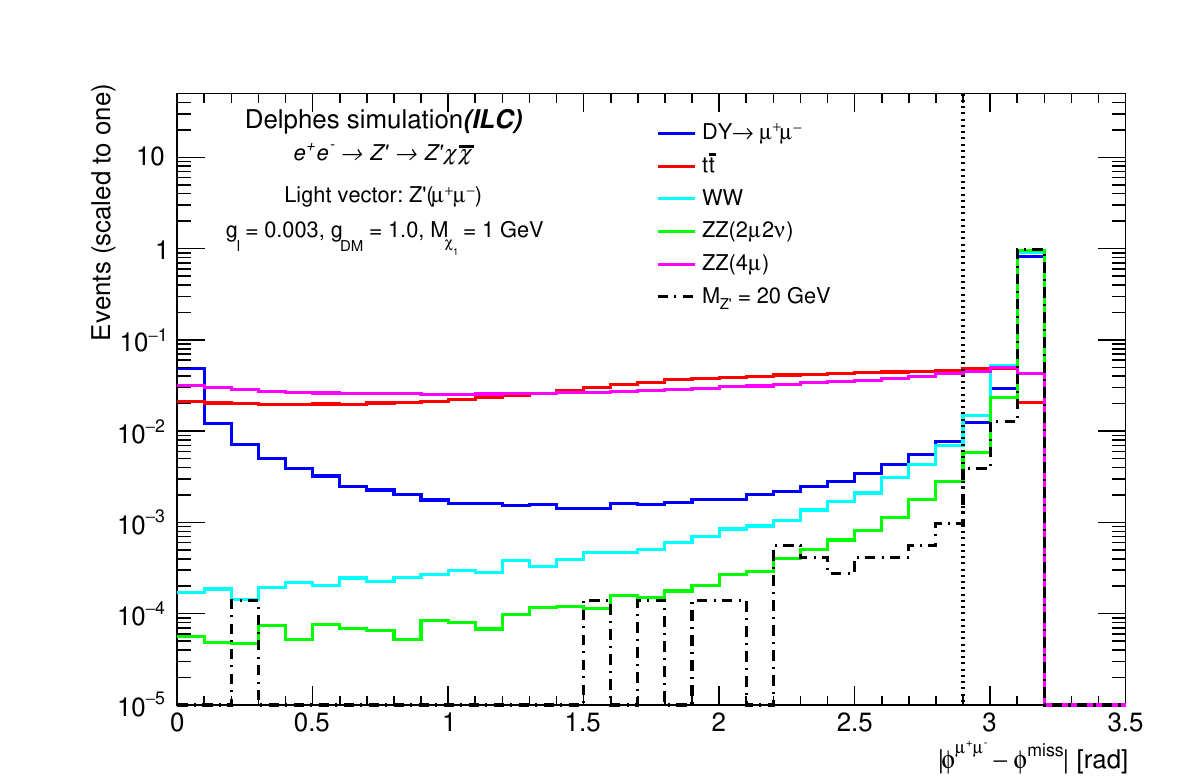}
}
\hspace{0mm}
%\centering
\subfigure[$\Delta R(\mu^{+}\mu^{-})$]{
  \includegraphics[width=92mm]{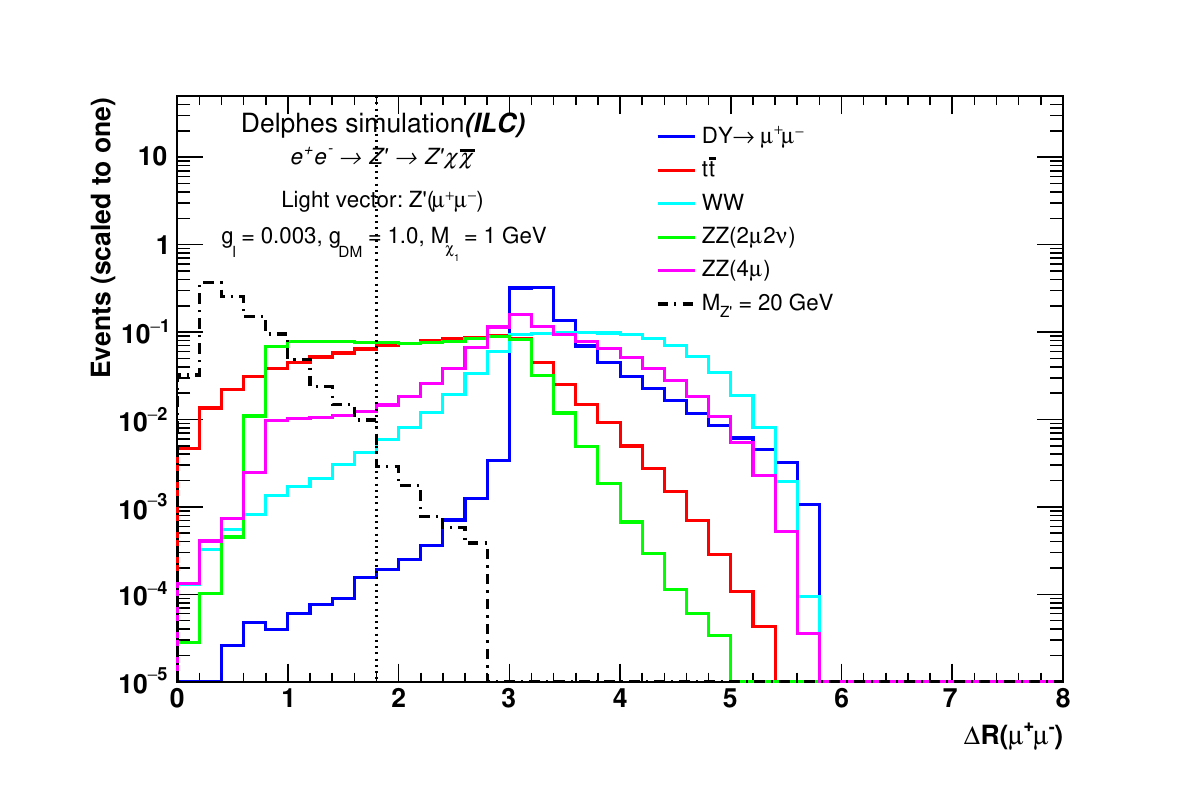}
}
\caption{The distributions of three variables for dimuon events, where each muon passes the low $p_T$ muon ID discussed in selection (i) in table \ref{section:AnSelection}. The three variables are $|E_{T}^{\mu^{+}\mu^{-}} - E_{T}^{miss}|/E_{T}^{\mu^{+}\mu^{-}}$ (a), $\Delta\phi_{\mu^{+}\mu^{-},\vec{E}_{T}^{miss}}$ (b), and $\Delta R(\mu^{+}\mu^{-})$ (c). The model corresponds to the LV scenario with $M_{Z^{\prime}}=20$ GeV and SM backgrounds. The histograms are normalized to unity to highlight qualitative features, and the vertical dashed lines correspond to the chosen cut value per each variable.}
\label{figure:fig70}
\end{figure}
%============================================================

In addition to the preliminary selection, extra tighter cuts have been applied. 
These tight cuts are based on four variables: the first variable is related to the invariant mass of the dimuon, at which we restricted the invariant mass of the dimuon to a small range around the mass of the neutral gauge boson Z$^{\prime}$, thus, it is required that $0.9 \times M_{Z^{\prime}} < M_{\mu^{+}\mu^{-}} < M_{Z^{\prime}} + 25$ as suggested in \cite{R1}. 
The second is the relative difference between the transverse energy of dimuon $(E_{T}^{\mu^{+}\mu^{-}})$ and the missing transverse energy $(E^{\text{miss}}_{T})$, it has been selected to be less than 0.4. 
(i.e. $|E_{T}^{\mu^{+}\mu^{-}} - E^{\text{miss}}_{T}|/E_{T}^{\mu^{+}\mu^{-}} < 0.4$).
The third one is $\Delta\phi_{\mu^{+}\mu^{-},\vec{E}^{\text{miss}}_{T}}$, which is defined as difference in the azimuth angle between the dimuon direction and the missing transverse energy direction (i.e. $\Delta\phi_{\mu^{+}\mu^{-},\vec{E}^{\text{miss}}_{T}} = |\phi^{\mu^{+}\mu^{-}}-~\phi^{miss}|$ ), it has been selected to be greater than 2.9 rad.
The fourth cut is the angular distance (or angular separation) between two opposite-sign muons ($\Delta R(\mu^{+}\mu^{-})$), it has to be less than 1.8.

The graphs, presented in figure \ref{figure:fig70}, illustrate the distributions of certain variables for the signal presentation of the simplified model relating to the light vector scenario. These variables are presented alongside SM backgrounds for dimuon events that pass selection (i). 
The first variable is denoted as $|E_{T}^{\mu^{+}\mu^{-}} - E_{T}^{\text{miss}}| / E_{T}^{\mu^{+}\mu^{-}}$, and its graph is shown in figure \ref{figure:fig70}(a). 
The second variable is denoted as $\Delta\phi_{\mu^{+}\mu^{-},\vec{E}_{T}^{\text{miss}}}$, and its graph is shown in figure \ref{figure:fig70}(b). 
The third variable is the angular distance ($\Delta R(\mu^{+}\mu^{-})$), and its graph is presented in figure \ref{figure:fig70}(c).
The model was generated with a neutral gauge boson mass of $M_{Z^{\prime}}$ = 20 GeV, and dark matter mass of $M_{\chi_1}$ = 1 GeV. These distributions are scaled to one. The vertical dashed lines in these figures correspond to the chosen cut value for each variable. 

These tight cuts are indicated in selection (ii) in table \ref{cuts}, they have significantly reduced the SM backgrounds.

%=============================================================
\section{Results}
\label{section:Results}
\begin{figure}%[h!]
\centering
  \resizebox*{9cm}{!}{\includegraphics{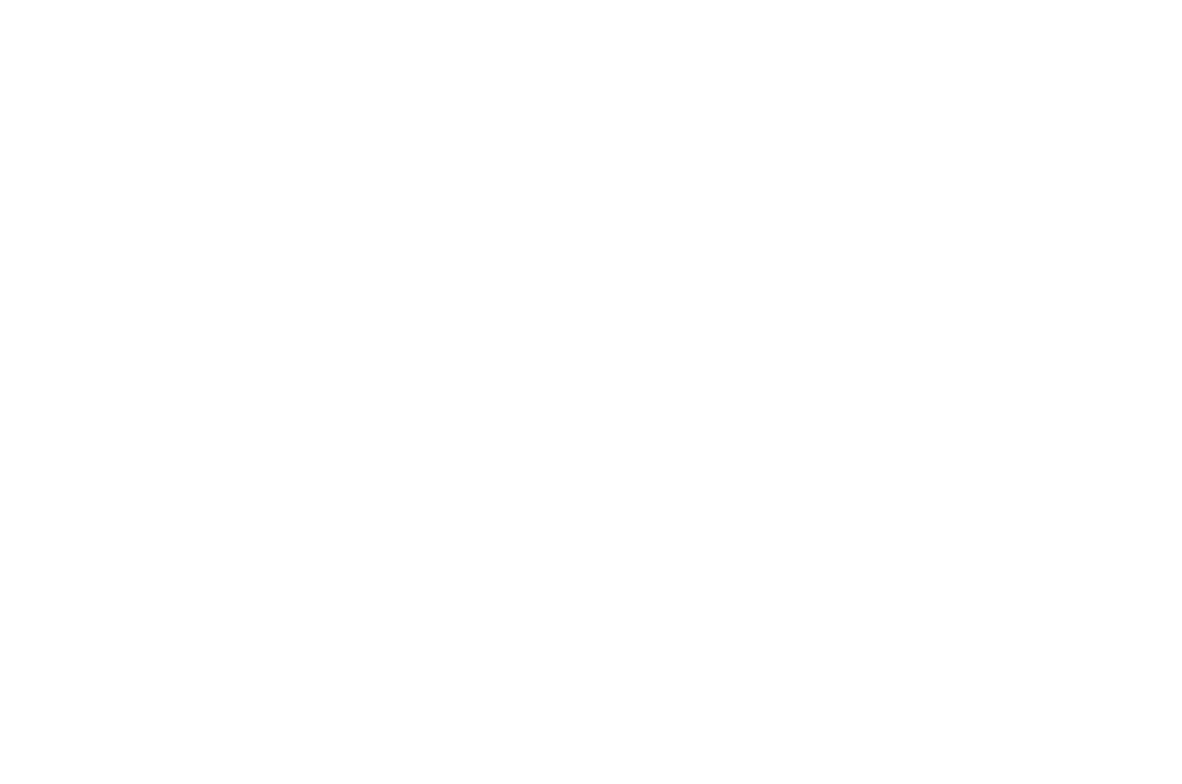}}
  \caption{The distribution of the missing transverse energy, after applying the final analysis selection(i+ii) listed in table \ref{cuts}, for the expected SM background and one signal benchmark corresponding to the LV with $M_{Z^{\prime}} = 20$ GeV is superimposed.}
  \label{figure:fig6}
\end{figure}

The shape-based analysis has been used based on the missing transverse energy distributions ($E^{\text{miss}}_{T}$), which are good discriminate variables since the signals distributions are characterized by relatively large $E^{\text{miss}}_{T}$ values compared to the SM backgrounds.
After applying the final event selection listed in table \ref{cuts}, the distribution of the missing transverse energy is illustrated in figure \ref{figure:fig6}. 
%%%%%%%%%%% plots step-2 %%%%%%%%%%%%%%%%%%%%%%%%%%%%%
%%%%%%%%%%% plots step-2 %%%%%%%%%%%%%%%%%%%%%%%%%%%%%
Table \ref{table:tab18} displays the results of the event selection process for both the SM backgrounds and the signal of the simplified LV scenario. The event selection process includes passing the analysis selection (i) and final selection (i+ii). The signal was generated with masses of a light gauge boson $M_{Z^{\prime}}$ of 20 GeV, and a dark matter mass $M_{\chi_{1}}$ of 1 GeV, corresponding to an integrated luminosity of 1000 fb$^{-1}$. The uncertainties in the results comprise both statistical and systematic components, which are summed in quadrature.

The significance of the signal over the background was calculated using the Asimov formula described in \cite{R2}.
\begin{equation}
     %S =  2 \times \big(\sqrt{N_s + N_b} - \sqrt{N_b}\big),
    S = \sqrt{2 \times \Big( (N_s + N_b) log\Big(1 + \frac{N_s}{N_b}\Big) - N_s\Big)},
    \label{sig:equ}
\end{equation}
where $N_s$ and $N_b$ are the number of signals and a total of SM background events passing the selections (i+ii) listed in table \ref{cuts}. 

\begin{figure}[h]
\centering
 \resizebox*{9.5cm}{!}{\includegraphics{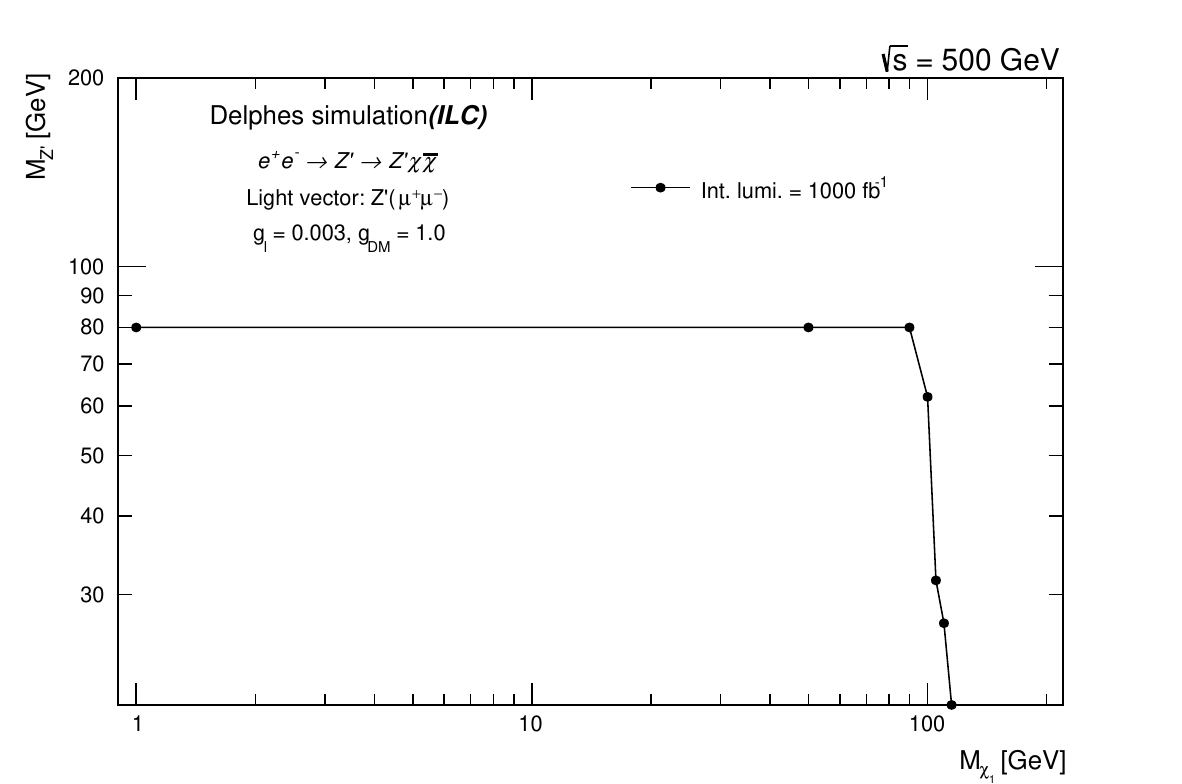}}
  \caption{$5\sigma$ detection region on the $M_{\chi_{1}}$ - $M_{Z^{\prime}}$ plane for 
  the $\chi \bar{\chi} Z^{\prime}$($Z^{\prime} \rightarrow \mu^{+}\mu^{-})$ production induced by the light vector scenario at the $\sqrt{s} = $ 500 GeV ILC with    integrated luminosity of $1000 ~\text{fb}^{-1}$.}
  \label{figure:significance2}
\end{figure}

In figure \ref{figure:significance2}, we depict the $5\sigma$ detection region (defined
as $S \geq 5$) on the $M_{\chi_{1}}$ - $M_{Z^{\prime}}$ plane by taking $\sqrt{s} = $ 500 GeV, integrated luminosity of $1000 ~\text{fb}^{-1}$ for the LV scenario, and fixing the values of the coupling constants $\texttt{g}_{l} = 0.003,~\texttt{g}_{DM} = 1.0$.
%%%%%%%%% plots for significance %%%%%%%%%%%%%%%%%%%%%%%%%%%%%
%%%%%%%%% plots for significance %%%%%%%%%%%%%%%%%%%%%%%%%%%%%
%\end{comment}
% ========== exclusion limits plot =======================
%%%%%%%%%%%%% plots step-1 %%%%%%%%%%%%%%%%%%%%%%%%%%%%%%%
\begin{figure}%[h!]
\centering
\subfigure[$M_{\chi_{1}} = 1$ GeV]{
  \includegraphics[width=92mm]{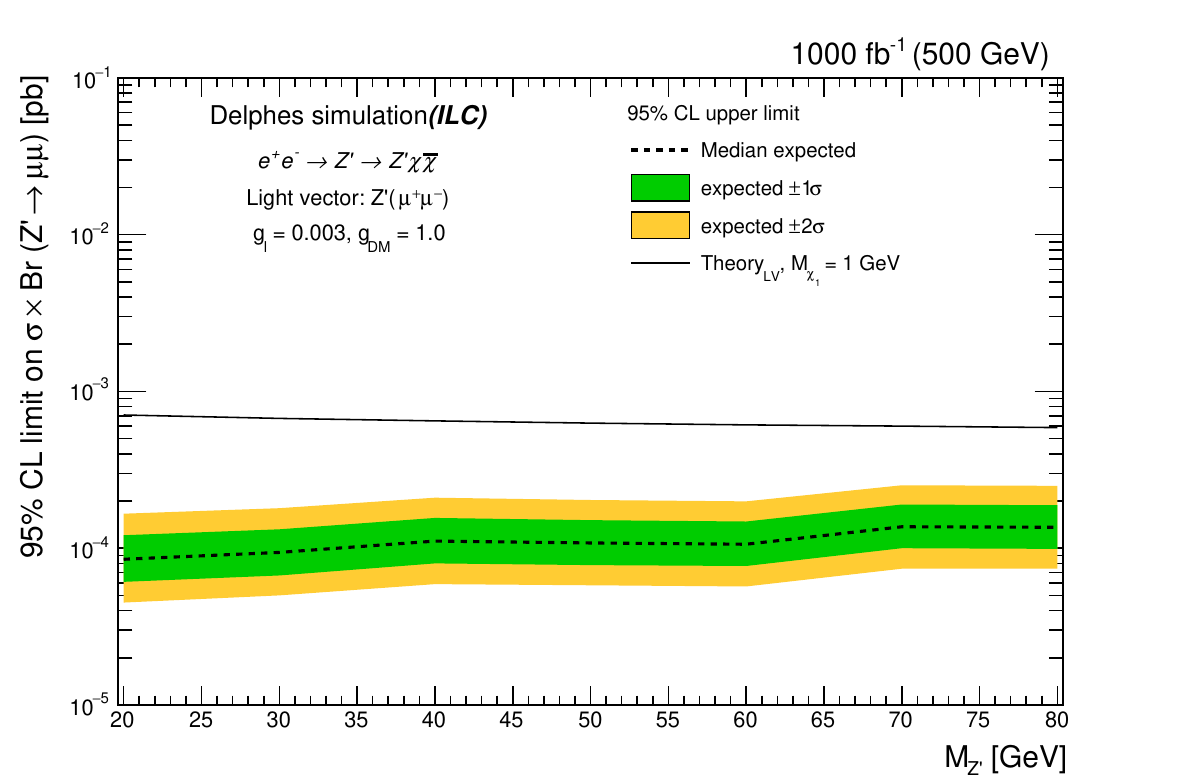}
}
\hspace{0mm}
%\centering
\subfigure[$M_{\chi_{1}} = 150$ GeV]{
  \includegraphics[width=92mm]{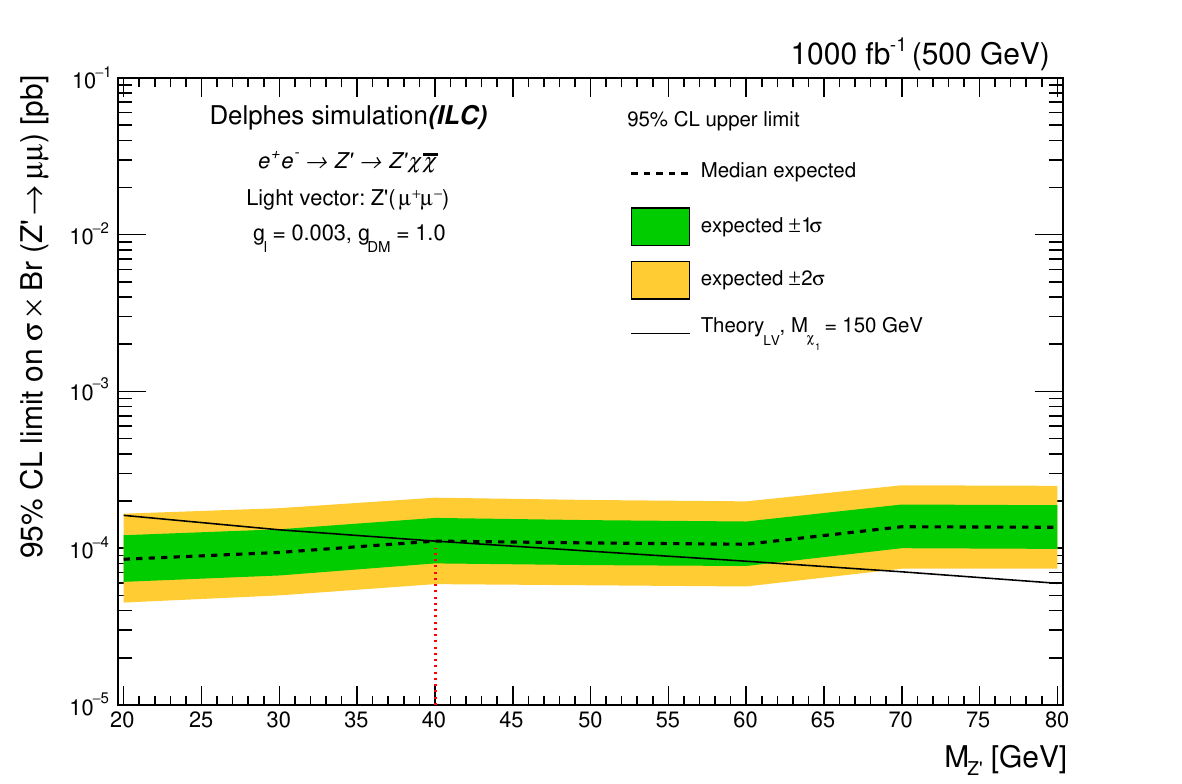}
}
\hspace{0mm}
%\centering
\subfigure[$M_{\chi_{1}} = 170$ GeV]{
  \includegraphics[width=92mm]{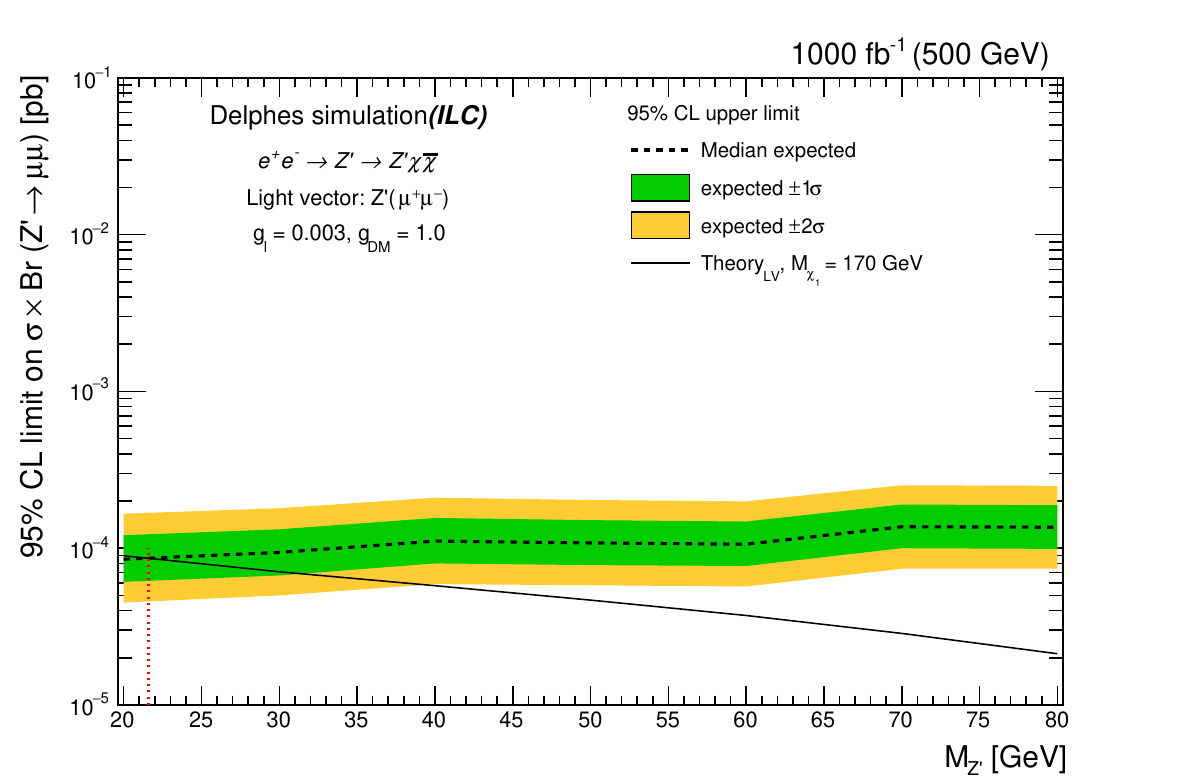}
}
\caption{95\% CL upper limits on the cross-section times the branching ratio (expected), as a function of the mediator's mass ($M_{Z^{\prime}}$) based on Mono-Z$^{\prime}$ model, with the muonic decay of the Z$^{\prime}$. The black line represents the light vector scenario with $M_{\chi_{1}} =$ 1 GeV (a), 150 GeV (b) and 170 GeV (c). The vertical dotted red line indicates the upper limit value.}
\label{figure:fig7}
\end{figure}
%%%%%%%%%%%%%%%%%%%%%%%%%%%%%%%%%%%%%%%%%%%%%%%%%%%%%%%%%%%

%%%%%%%%%%%%%%% no of events %%%%%%%%%%%%%%%%%%%%%
\begin{comment}
\begin{table}%[h]
\centering
\label{ tab-marks }
\begin {tabular} {|l|c|}
\hline
Process & No. of events \\
\hline
\hline
$\text{DY} \rightarrow \mu^{+}\mu^{-}$ & $388.3 \pm 43.5$  \\
\hline
$\text{t}\bar{\text{t}} \rightarrow \mu^{+}\mu^{-} + 2b + 2\nu$   & $111.8 \pm 15.4$\\
\hline
$\text{WW} \rightarrow \mu^{+}\mu^{-} + 2\nu$ & $118.4 \pm 16.1$ \\
\hline
$\text{ZZ} \rightarrow \mu^{+}\mu^{-} + 2\nu$ & $0.4 \pm 0.7$ \\
\hline
$\text{ZZ} \rightarrow 4\mu$   & $0.5 \pm 0.7$ \\
\hline
Sum Bkgs & $619.4 \pm 66.8$  \\
\hline
\hline
%Data & 61 \\
%\hline
%\hline
Signal of LV scenario & $11366.5 \pm 1141.6$   \\
(at $M_{Z^{\prime}}$ = 50 GeV and $M_{\chi_{1}} = 1$ GeV) &  \\
\hline
\end {tabular}
\caption{The number of events, satisfying the criteria of the final event selection(i+ii) listed in table \ref{cuts}, are illustrated for each SM background and the the simplified model in the Mono-Z$^{\prime}$ portal with coupling constants $g_{DM} = 1.0$,  $g_{l} = 0.003$ and $M_{\chi_{1}} = 1$ GeV, corresponding to a 1000 fb$^{-1}$ integrated luminosity at the ILC with $\sqrt{s} = 500$ GeV. 
The total  uncertainty, including the statistical and systematic components, is indicated.}
\label{table:tab8}
\end{table}
\end{comment}
%===================================================
%%%%%%%%%%%%%%% no of events %%%%%%%%%%%%%%%%%%%%%
\begin{table*}
    \centering
    \begin{tabular}{|c|c|c|}
\hline
Process & No. of events passing (i)& No. of events passing (i+ii)\\
\hline
\hline
$\text{DY} \rightarrow \mu^{+}\mu^{-}$ & $875755.7 \pm 87580.6$ & $0 \pm 0$\\
\hline
$\text{t}\bar{\text{t}} \rightarrow \mu^{+}\mu^{-} + 2b + 2\nu$ & $6292.7 \pm 634.3$ & $54.9 \pm 9.2$\\
\hline
$\text{WW} \rightarrow \mu^{+}\mu^{-} + 2\nu$ & $136028.5 \pm 13607.8$ & $732.3 \pm 78.1$ \\
\hline
$\text{ZZ} \rightarrow \mu^{+}\mu^{-} + 2\nu$ & $2515.3 \pm 256.5$ & $11.9 \pm 3.7$ \\
\hline
$\text{ZZ} \rightarrow 4\mu$ & $462.5 \pm 51.0$ & $0.1 +/-   0.4 $ \\
\hline
Sum Bkgs & $1021054.6 \pm 102110.5$ & $799.2 \pm 84.8$ \\
\hline
\hline
Signal of LV scenario & $366.3 \pm 41.3$ & $345.0 \pm 39.2$  \\
(at $M_{Z^{\prime}}$ = 20 GeV and $M_{\chi_{1}} = 1$ GeV) &&  \\
\hline
\end {tabular}
\caption{The table displays the number of events that passed the pre-selection (middle column) and the full selection (right column) criteria, as obtained from simulations for backgrounds and a signal. The simulations were performed with a luminosity of 1000 fb$^{-1}$ at $\sqrt{s} = 500$ GeV. 
The signal sample corresponds to the simplified-model scenario LV with $M_{Z^{\prime}}$ = 20 GeV, $M_{\chi_{1}} = 1$ GeV, $g_{DM} = 1.0$ and $g_{l} = 0.003$. 
The total uncertainties, including both the statistical and systematic components, have been taken into account for the simulated signal and background samples.}
\label{table:tab18}
\end{table*}
%=============================================================

%To make a statistical interpretation for our results, we performed a statistical test based on the profile likelihood method, with the use of the modified frequentist
%construction CLs [38, 39] used in the asymptotic approximation [40] to derive exclusion limits on the product of signal cross sections and branching fraction Br(Z′ → µµ)
%at 95% confidence level.

For the purpose of analyzing our results statistically, we utilized the profile likelihood method and performed a statistical test.
We used the modified frequentist construction CLs \cite{R58, R59}, which is based on the asymptotic approximation \cite{R2}, to derive exclusion limits on the product of signal cross sections and the branching fraction Br($Z^{\prime}$ $\rightarrow \mu\mu$) at a 95\% confidence level.

In the Mono-Z$^{\prime}$ model, the 95\% upper limit on the cross-section times the branching ratio for the LV simplified scenario is shown in Figure \ref{figure:fig7}. 
The result is presented for the muonic decay of the Z$^{\prime}$ and with coupling constant values of $\texttt{g}_{l} = 0.003$ and $\texttt{g}_{DM} = 1.0$, for an integrated luminosity of 1000 fb$^{-1}$. 
The limits are illustrated for dark matter mass $M_{\chi_{1}}$ values of 1 GeV (Figure \ref{figure:fig7}.a), 150 GeV (Figure \ref{figure:fig7}.b), and 170 GeV (Figure \ref{figure:fig7}.c), which are represented by the black solid curves. 
The vertical dotted red line indicates the upper limit value.

\begin{figure}%[h!]
\centering
  \resizebox*{9.0cm}{!}{\includegraphics{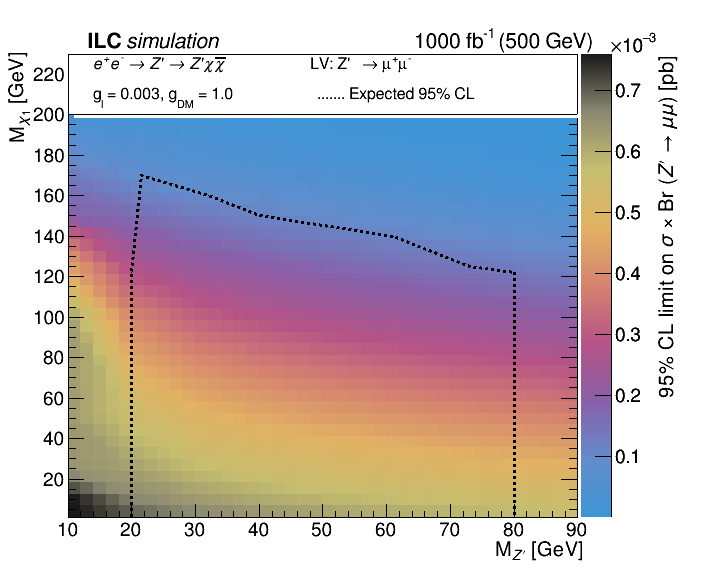}}
  \caption{The 95\% CL upper limits on the product of the cross-section and branching fraction from the inclusive search, for variations of pairs of the LV scenario parameters ($M_{Z^{\prime}}$ and $M_{\chi_{1}}$). The filled region indicates the upper limit. The dotted black curve indicates the expected exclusions for the nominal Z$^{\prime}$ cross-section.}
  \label{figure:fig8}
\end{figure}

Figure \ref{figure:fig8} shows the limit on the cross-sections times the branching ratios for the muonic decay channel of the Z$^{\prime}$ boson as functions of the mediator's mass ($M_{Z^{\prime}}$) and the mass of the dark matter ($M_{\chi_{1}}$). The region inside the contour is excluded for the benchmark scenario where $\texttt{g}_{l} = 0.003$ and $\texttt{g}_{DM} = 1.0$. We used an integrated luminosity of 1000 fb$^{-1}$.
This limit shows that the invariant mass range from 20 to 80 GeV is excluded for $M_{\chi_{1}} \in [1, 122]$ GeV.
%=========================================

%================================================================
%================================================================
%\newpage
%================================================================
\section{Summary}
\label{section:Summary}
The ILC electron-positron collider is an optimal machine for the possible detection of particles from BSM since it delivers a clean signature of unknown particles as dark matter, extra neutral gauge bosons, and Kaluza-Klein excitation concerning the QCD background.

In this view, we have studied the effects of a simplified-model scenario, which is known as light vector (LV), via dark matter pair production associated with a Z$^{\prime}$ boson at the ILC. 
The LV signal samples have been simulated based on electron-positron collisions corresponding to the foreseen ILC RUN I with 500 GeV center of mass energy, for an integrated luminosity of 1000 fb$^{-1}$. 
Results from the muonic decay mode of Z$^{\prime}$ are discussed, with fixing the values of the coupling constants to be $g_{DM} = 1.0$, $g_{l} = 0.003$.
Given that the polarized degrees of electron and positron beams are $P_{e^{-}} = 0.8$, $P_{e^{+}} = -0.3$ respectively at the ILC.

If this signal is not observed at the ILC, we set upper limits on the masses of Z$^{\prime}$ and dark matter $(\chi_{1})$ at the 95\% CL for the charged muonic channel decay of Z$^{\prime}$. 
Limits have been set for light vector scenario with $\texttt{g}_{l} = 0.003$ and $\texttt{g}_{DM} = 1.0$, excluding the invariant mass range from 20 to 80 GeV for
$M_{\chi_{1}} \in [1, 122]$ GeV, nevertheless excluding $M_{\chi_{1}}$ = 170 GeV at $M_{Z^{\prime}}$ = 21.5 GeV.

%================================================================
%================================================================
\begin{acknowledgments}
The author of this paper would like to thank Tongyan Lin, co-author of \cite{R1} for providing us with the UFO model files, helping us with generating the signal events, and cross-checking the results.
This paper is based on works supported by the Science, Technology, and Innovation Funding Authority (STDF) under grant number 48289.
\end{acknowledgments}
%\nocite{*}

%\newpage
%========= References ===============================================

\end{document}